\documentclass[useAMS,usenatbib]{mn2e}

\usepackage{graphicx}
\usepackage{url}
\usepackage{natbib}
\usepackage{my_aas_macros}
\usepackage{caption}
\usepackage{array}
\usepackage{multicol}
\usepackage{amsmath}

\title[Indications of stellar prominence oscillations]{Indications of stellar prominence oscillations 
on fast rotating stars: the cases of HK~Aqr and PZ~Tel\thanks{Based on observations made with ESO Telescopes at the LaSilla Observatory under programme ID 089.D-0709(A).}}
\author[M. Leitzinger et al.]{M. Leitzinger$^{1}$\thanks{E-mail: martin.leitzinger@uni-graz.at}, P. Odert$^{2,1}$, T.V. Zaqarashvili$^{1,2,3}$, R. Greimel$^{1}$, \newauthor A. Hanslmeier$^{1}$, H. Lammer$^{2}$
\\
$^{1}$Institute of Physics, Department for Geophysics, Astrophysics, and Meteorology, NAWI Graz, Universit\"atsplatz 5, 8010 Graz, Austria\\
$^{2}$Space Research Institute, Austrian Academy of Sciences, Schmiedlstra\ss{}e 6, 8042, Graz, Austria\\
$^{3}$Abastumani Astrophysical Observatory, Ilia State University, Tbilisi, Georgia\\
}
\begin{document}

\date{Accepted 2016 August 1. Received 2016 July 8; in original form 2015 December 2}

\pagerange{\pageref{firstpage}--\pageref{lastpage}} \pubyear{2002}

\maketitle

\label{firstpage}

\begin{abstract}
We present the analysis of six nights of spectroscopic monitoring of two young and fast rotating late-type stars, namely the dMe star HK~Aqr and the dG/dK star PZ~Tel. On both stars we detect absorption features reminiscent of signatures of co-rotating cool clouds or prominences visible in H$\alpha$. Several prominences on HK~Aqr show periodic variability in the prominence tracks which follow a sinusoidal motion (indication of prominence oscillations). On PZ~Tel we could not find any periodic variability in the prominence tracks. By fitting sinusoidal functions to the prominence tracks we derive amplitudes and periods  which are similar to those of large amplitude oscillations seen in solar prominences. In one specific event we also derive a periodic variation of the prominence track in the H$\beta$ spectral line which shows an anti-phase variation with the one derived for the H$\alpha$ spectral line. Using these parameters and estimated mass density of a prominence on HK~Aqr we derive a minimum magnetic field strength of $\sim$2~G. 
The relatively low strength of the magnetic field is explained by the large height of this stellar prominence ($\ge$ 0.67 stellar radii above the surface).
\\
\end{abstract}


\section{Introduction}
\label{Intro}
Prominences are manifestations of solar/stellar coronal magnetic fields. Coronal magnetic field supports cool dense prominence plasma against gravity, which may be kept for several rotations (quiescent prominences) or may be ejected because of disturbances in the stellar plasma causing the plasma to accelerate very fast (eruptive prominences). Prominences and filaments refer to the same phenomenon with the difference that prominences are seen against the dark background at the solar/stellar limb in emission and filaments are seen against the solar/stellar disk in absorption. We will use the term ``prominence'' throughout the paper. Eruptive prominences often lead to Coronal Mass Ejections (CMEs) which propagate through the solar/stellar corona and interplanetary space where they normally decelerate and dissolve into the solar/stellar wind. There exists a good correlation of solar CMEs and eruptive prominences \citep{Hori2002, Gopalswamy2003} which seems to be not surprising as the erupting filament represents the CME core. On the Sun, prominences/filaments and CMEs have been studied since decades and data covering a complete solar cycle allow statistical analyses \citep{Gopalswamy2003}.\\
Solar quiescent prominences have lifetimes between 1 and 300 days and a mean height of 2.6$\times$10$^{4}$~km \citep{Wang2010}. The masses of solar prominences are not easily determined from observations, therefore only a few measurements of selected prominences exist. Using typical values for hydrogen column density and volume of solar prominences, masses in the range of 5$\times$10$^{12}$-10$^{15}$g are derived \citep{Labrosse2010}.\\
On the Sun, we know that prominences show oscillations, both longitudinal and transverse. Oscillating prominences may lead to eruptions sometimes. Large amplitude oscillations are oscillations which are triggered by Moreton or EIT waves, flares, or are related to the eruptive phase of filaments. The prominence oscillations are divided into small-amplitude (2-3~km~s$^{-1}$) and large-amplitude ($>$20~km~s$^{-1}$) oscillations \citep[see][]{Arregui2012}. Periods of large amplitude oscillations are in the range of 6-150~min \citep{Tripathi2009} while small amplitude oscillations have periods up to 90~minutes. But even periods below one minute and above several hours are reported for small amplitude oscillations \citep{Arregui2012}. The so-called winking filaments are manifestations of large-amplitude oscillations. Solar H$\alpha$ filters have a certain narrow bandpass. The large-amplitude oscillation periodically shifts the H$\alpha$ line outside of the filter bandpass making the filament ``invisible''. Therefore they were named ''winking``.\\
Stellar prominences have been detected for the first time by \citet{CollierCameron1989a} on the rapidly \citep[0.52~d][]{Innis1988} rotating K-star AB Doradus as transient absorption features in H$\alpha$. Prominences can only be detected in rotationally broadened H$\alpha$ line profiles while the absorption transients travel over the H$\alpha$ line profile. In a rotationally unbroadened H$\alpha$ line profile such absorption transients cannot be detected. Up to now, prominences have been detected on a number of fast rotators such as AB Doradus \citep{CollierCameron1989a, CollierCameron1989b}, BO~Mic \citep{Jeffries1993, Dunstone2006a}, HK~Aqr \citep{Byrne1996}, PZ~Tel \citep{Barnes2000}, G dwarf stars in the $\alpha$ Persei cluster \citep{CollierCameron1992}, the post T Tauri star RX J1508.6-4423 \citep{Donati2000} and EY~Dra \citep{Eibe1998}. The prominences of the stars reach heights of several stellar radii, even  above the co-rotation radius. Masses have been determined for AB~Dor in the range of 2-6$\times$10$^{17}$g \citep{CollierCameron1990} and for Speedy Mic in the range of 0.5-2.3$\times$10$^{17}$g \citep{Dunstone2006b}. Moreover, prominences have been also detected on binary systems as absorptions in optical spectra \citep{Parsons2011} and dips in light-curves lasting for several rotation periods \citep{Irawati2015}. Oscillations in stellar prominences derived from absorption transients have so far not been detected. \\
Within this study we aim for the detection of prominence oscillations which can be used to estimate the magnetic field strength in stellar prominences. This method is a further development of solar coronal seismology, which uses observed oscillations to determine plasma parameters \citep{Nakariakov2001}. Moreover, stars hosting prominences should also be potential candidates for the detection of prominence eruptions as we know from the Sun that the core of a CME is the prominence/filament. As targets for this study we have chosen two stars of different spectral type known to host prominences. The first one is the fast rotating \citep[0.94d, ][]{Innis1990} solar analog pre-main sequence star PZ~Telescopii with an age of $\sim$12~Myr \citep{Zuckerman2001}, a spectral type of G6.5 to K8 \citep{Schmidt2014, Messina2010}, and a prominence detection reported in \citet{Barnes2000}. The distance to PZ~Tel is 49.7~pc (derived from the Hipparcos parallax) and its $v$sin$i$ is $\sim$ 73~km~s$^{-1}$ \citep{Jenkins2012}. The second star is the fast rotating \citep[0.43d, ][]{Young1990}, young \citep[200~Myr, ][]{Montes2001} and active dM1.5e \citep{Barnes2004} star HK~Aquarii with a prominence detection reported in \citet{Byrne1996}. The distance to HK~Aqr is 22.3~pc and its $v$sin$i$ is 70~km~s$^{-1}$ \citep{Young1990}. HK~Aqr shows H$\alpha$ in emission and PZ~Tel in absorption.

\section[]{Observations}
We obtained 6 nights of observations at the European Southern Observatory (ESO) with the Fiber-fed Extended Range Optical Spectrograph (FEROS) mounted on the 2.2m telescope of the Max Planck Gesellschaft (MPG) on La Silla. 
The spectral resolution of FEROS (R=48000) is sufficient to search for prominence oscillations. The velocity resolution corresponds to $\sim$6~km~s$^{-1}$, sufficient to detect the stellar analogue of solar large amplitude oscillations. The total number of spectra for PZ~Tel was 230  with an exposure time of 5~minutes each resulting in a total on-source time of 1150 minutes. The total number of spectra for HK~Aqr was 110  with an exposure time of 10~minutes each resulting in a total on-source time of 1100 minutes.\\
The observations were scheduled from the 7$^{th}$ to the 13$^{th}$ of August, 2012. We lost one half-night due to technical problems, therefore no observations for HK~Aqr were executed on the 10$^{th}$ of August. For details on the observations see Table~\ref{tabobs}. For an inspection of the quality of the data see Fig.~\ref{allspec08}, Fig.~\ref{allspec11}, Fig.~\ref{allspec11beta}, and Fig.~\ref{allspec12} in the Appendix.\\
\begin{table}
\centering
   \caption{Observational details of the ESO FEROS data used in the study. The S/N values refer to mean values for each night calculated around H$\alpha$.}
   \label{tabobs}
     \begin{tabular}{ccc}
  \hline
      date     &       no. of spectra     &      S/N           \\
               &            HK/PZ         &       HK/PZ        \\    
\hline      
    07/08/2012  &            10/49         &      46/74        \\   
    08/08/2012  &            17/60         &      38/98        \\
    09/08/2012  &            14/50         &      77/91        \\
    10/08/2012  &             -/30         &      -/109        \\
    11/08/2012  &            38/30         &      61/90        \\
    12/08/2012  &            31/10         &      55/106       \\
      \hline
\end{tabular}
\end{table} 
\subsection{Data reduction and preparation}
\label{dataprep}
The FEROS Echelle spectra have been reduced with the Image Reduction and Analysis Facility (IRAF) V2.16. The standard reduction steps were BIAS and Flat Field correction, cosmic ray removal, and wavelength calibration (ThArNe frames). The data have been normalized for analysis. 
To make prominence structures in H$\alpha$ visible we apply a common method, which is the subtraction of individual spectra by an average spectrum generated from the whole

\begin{figure*}
\begin{center}
\vspace*{0cm}
\includegraphics[width=6.8cm]{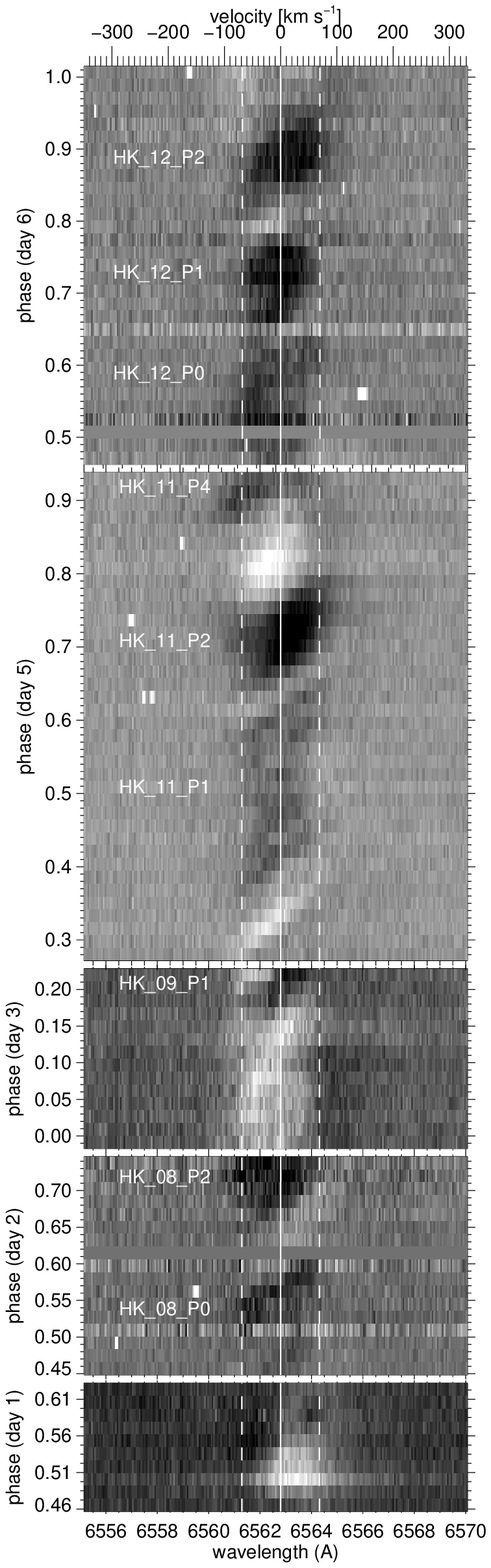}
\includegraphics[width=6.8cm]{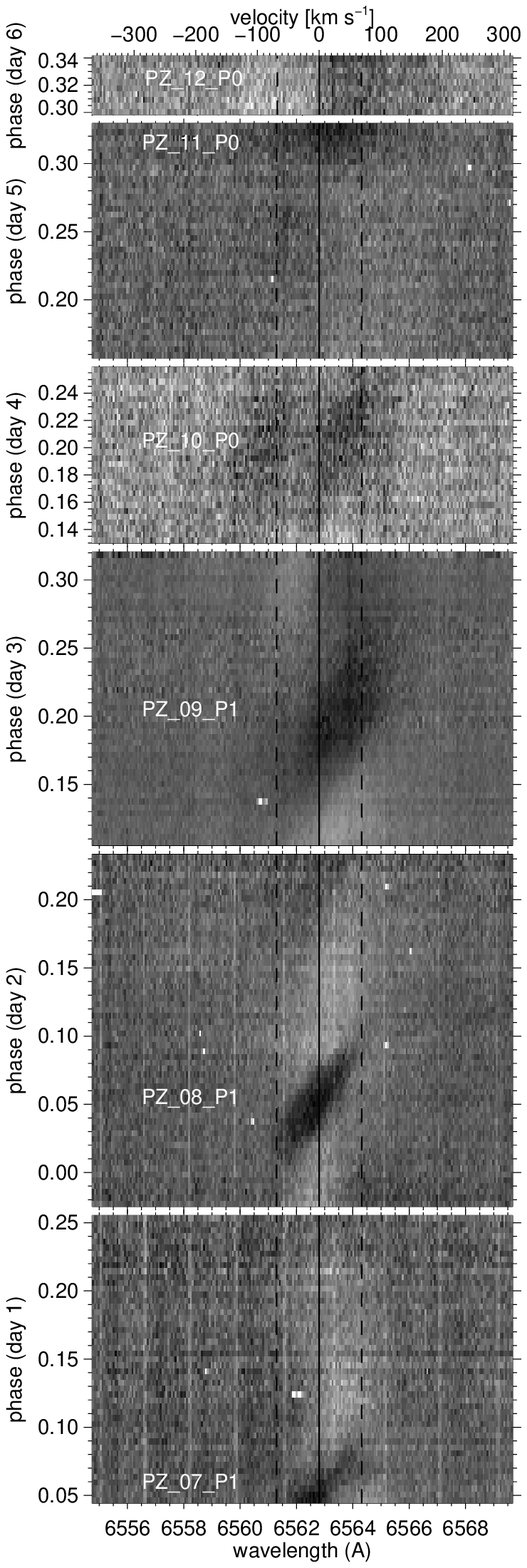}
 \caption{H$\alpha$ dynamic residual spectra of HK~Aqr (left panel) and PZ~Tel (right panel) starting from 07-08-2012 lowest panel to 12-08-2012 top panel (night of 10-08-2012 is missing as no observations of HK~Aqr were possible due to technical problems). The intensity in the images is grey scaled (black refers to minimum values whereas white corresponds to maximum values). One can clearly see drifting structures over the H$\alpha$ profile, both black and white. The prominences are named according to Table~\ref{Tab1}. The solid vertical lines mark the center of H$\alpha$ and the vertical dashed lines mark the $v$sin$i$ value of the corresponding star.}
\label{dynspecred}
 \end{center}
 \end{figure*}
\begin{figure}
\begin{center}
\vspace*{0cm}
\includegraphics[width=8.0cm]{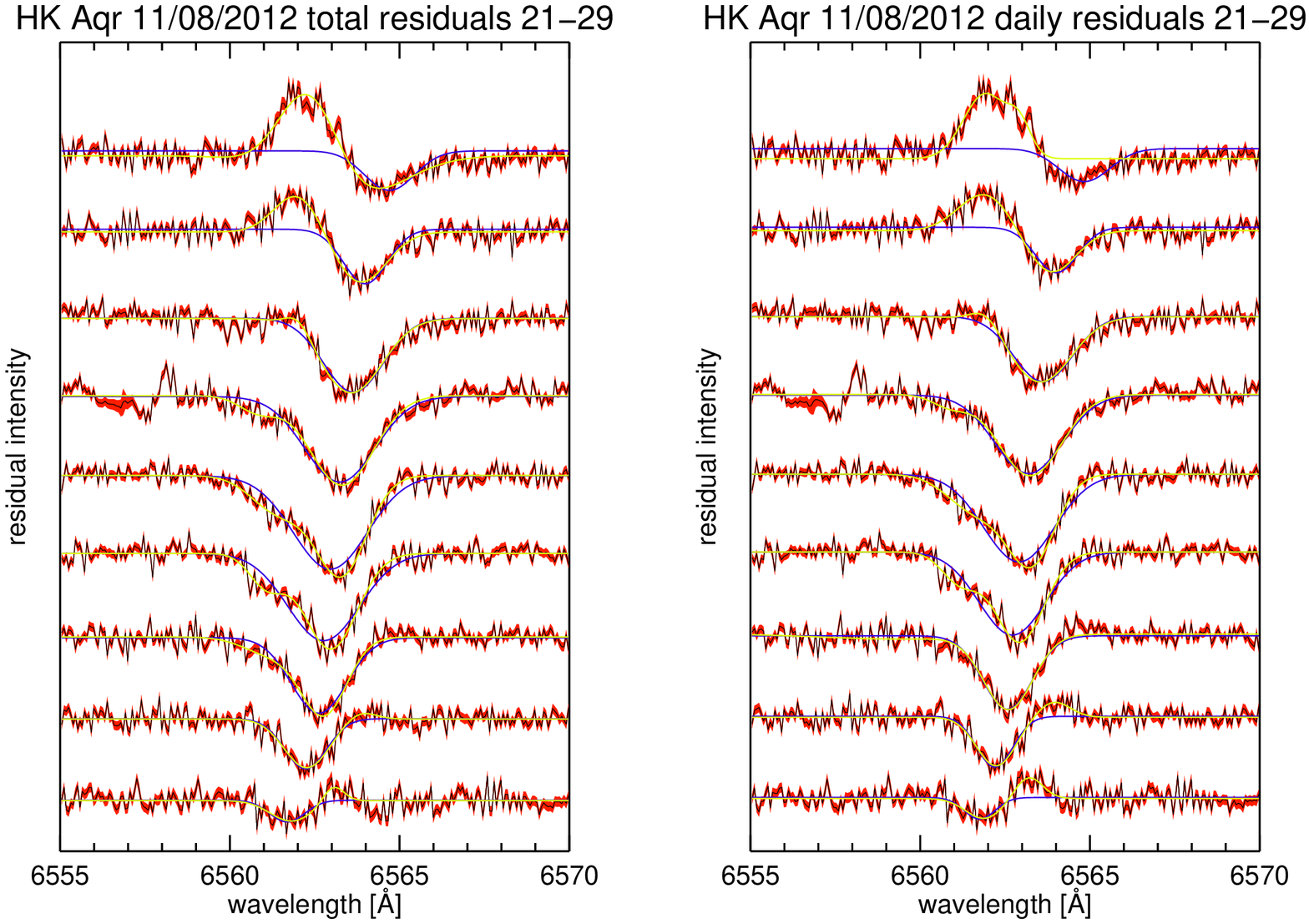}
\includegraphics[width=8.0cm]{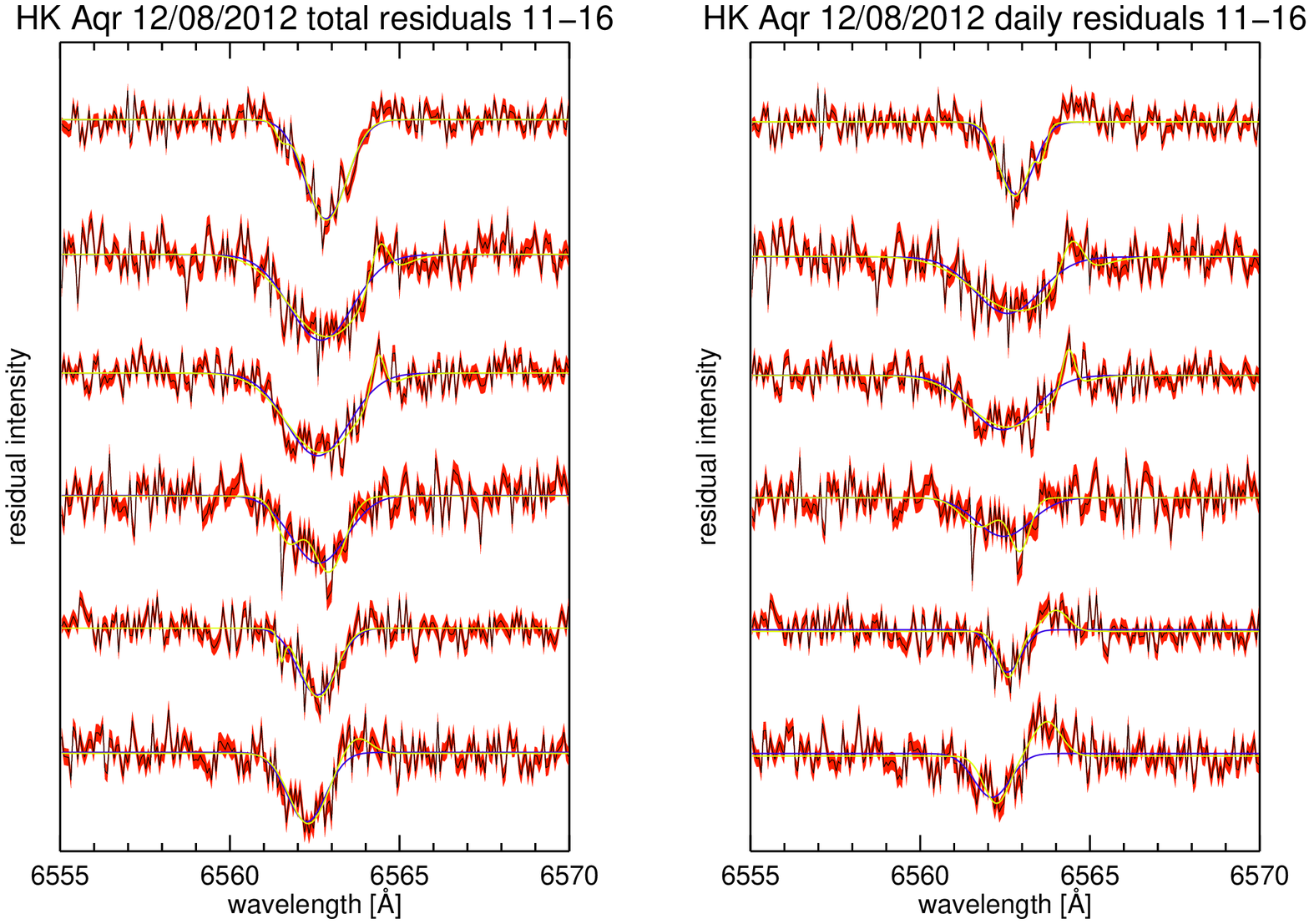}
 \caption{Upper and lower panels: Sequence of residuals (left panels: derived from total average subtraction; right panels: derived from daily average subtraction) of two prominences (HK\_11\_P2 and HK\_12\_P1) where we detected oscillatory motion. Overplotted are single (blue solid line) and double gaussian (yellow solid line) fits. The red colored area corresponds to the 1-$\sigma$ uncertainties per data point.}
\label{showspec}
 \end{center}
 \end{figure}

 
 
\noindent time series. The residual spectra are then plotted as a dynamic spectrum (wavelength vs. time) which is shown in Fig.~\ref{dynspecred} for HK~Aqr and PZ~Tel.\\
To determine the center positions of the prominence tracks we fitted the residual spectra, which we build using total and daily averages, with single and double gaussians (see section~\ref{stability}, Fig.~\ref{showspec}, Fig.~\ref{osco}). In this way we can deduce the central position of each absorption feature as it moves across the H$\alpha$ profile.\\ 
\citet{Dunstone2006a} investigated prominence systems on the ultra-fast rotator Speedy Mic. By generating dynamic spectra of their time series it was obvious that many absorption structures were visible already from the spectral time series without dividing or subtracting a mean profile. These authors avoided the above described simple method to obtain residual spectra because of the many strong and variable absorption signatures in H$\alpha$. Therefore these authors applied a gaussian weighted running mean in the temporal direction to obtain residual spectra. The dynamic spectra of HK~Aqr show only subtle absorption features, therefore it is possible to find a reference level which is not influenced by the absorptions. 

\noindent The same is true for PZ~Tel. Moreover, the application of a gaussian weighted running mean requires a longer spectral time series to achieve good results. As our observation strategy was directed to the observation of two stars during the same night the number of continuous spectra is therefore small. The application of an average spectrum as a reference level of H$\alpha$ clearly shows distinct absorption features with adequate contrast (cf. Fig.~\ref{showspec}). Therefore we decided to use this technique instead of an unsharp masking.
To evaluate the quality of the data, the significance of the absorption features in the residual spectra caused by the co-rotating prominences, and the quality of single and double gaussian fitting, we show in Fig.~\ref{showspec} two sequences of residual spectra 
of two prominences where we have detected oscillatory motion (see section~\ref{promosc}) of the center positions (HK\_11\_P2 and HK\_12\_P1). The red shaded area in the plots corresponds to the 1-$\sigma$ error per data point.

\section{Results}
\label{Results}
\subsection{Prominences on HK~Aqr and PZ~Tel}
\label{promi}
Fig.~\ref{dynspecred} shows dynamic residual spectra obtained from H$\alpha$ spectroscopic observations of HK~Aqr (left panel) and PZ~Tel (right panel) for each night of observation. Prominence tracks caused by co-rotating prominences can be identified as drifting structures from blue to red. One can see structures of different slopes, indicating different heights and/or occurrence on different stellar latitudes, as well as structures seen in absorption and emission. The majority of the structures is seen in absorption. The nature of the tracks seen in emission might be related to stellar activity such as plage regions. Prominences, seen off-disk in emission are visible as subtle emission features beyond the stellar $v$sin$i$ value. 
\citet{Dunstone2006b} nicely show prominences seen in emission above the stellar limb on the ultra-fast rotator Speedy Mic. Our data have a S/N too low to allow a detection of prominences in emission when they rotate off the disk. \\
We detect in total 8 prominence structures on HK~Aqr and 6 on PZ~Tel, all of them seen in absorption. In the spectroscopic time series of HK~Aqr of the first night there is one drifting structure seen in emission which clearly shows an enhancement during the first few spectra indicating a flare. In Table~\ref{Tab1} we present the prominence parameters of the detected structures. Those are height and projected area of the prominences, as well as prominence oscillation parameters, amplitude and period (see sections~\ref{Resprom}, ~\ref{promosc}, and \ref{promarea}).\\

\subsubsection{Prominences in H$\beta$}
\label{Hbetaproms}
As the FEROS spectra cover a wavelength range from 3500-9200\AA{} we are able to look for signatures of prominences also in other spectral lines. FEROS covers several Balmer lines but the S/N decreases towards shorter wavelengths due to reduced emission of the star and the fact that the sensitivity of the detector is lower at blue wavelengths. Because H$\beta$ is the second strongest line of the Balmer sequence  we have chosen it as a second spectral line to look for signatures of prominences and neglect H$\gamma$, H$\delta$, and further Balmer lines. Prominences are also visible in the CaII H+K lines but the S/N is much lower than at H$\alpha $ in our observations. 
Not all prominences which we have detected on HK~Aqr in H$\alpha$ can be seen in H$\beta$. Especially two events can be reproduced very well in H$\beta$, which are HK\_11\_P2 and HK\_12\_P1.
When detecting prominence signatures in H$\beta$ we can deduce the residual H$\beta$ flux and can compare it to the residual H$\alpha$ flux. By building a H$\alpha$/H$\beta$ flux ratio we are able to infer, if the prominence material is optically thick. If it is optically thick, then we may derive the area on the star which is obscured by the prominence (cf. section~\ref{promarea}) as well as its mass. The analysis of the H$\beta$ prominences has revealed that the linear fits to the prominences tracks deviate from the H$\alpha$ fits which is surprising, as we would have expected a similar slope because the signatures detected in both spectral lines originate from the same prominence. 


\subsection{H$\alpha$ and H$\beta$ light-curves}
\label{lightcurve_text}
\begin{figure}
\begin{center}
\vspace*{0cm}
\includegraphics[width=8.0cm]{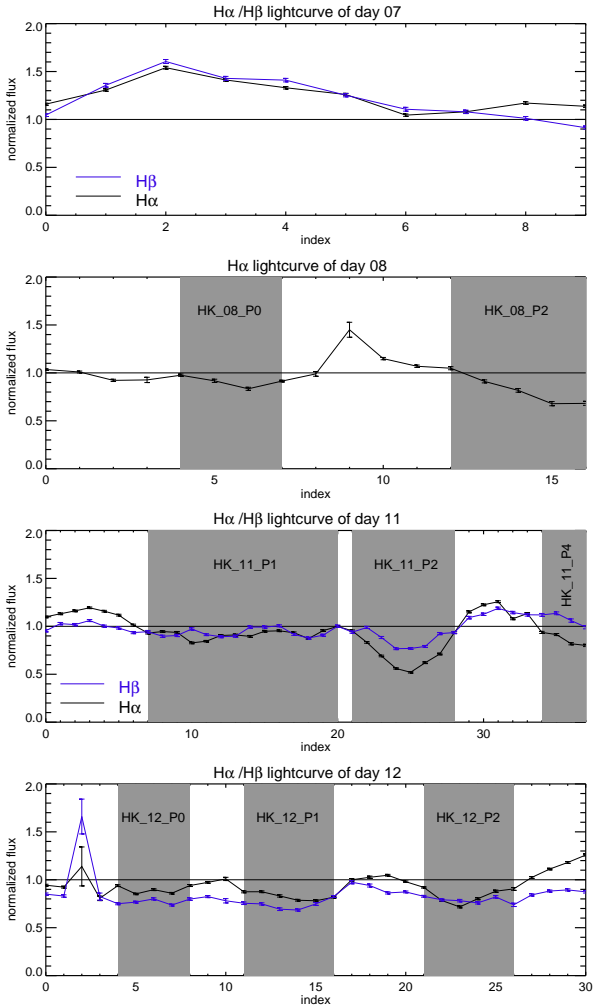}
 \caption{H$\alpha$ and H$\beta$ light-curves of the first night (upper panel) when a flare was detected and of nights (remaining panels) when we have detected prominences on HK~Aqr. The grey shaded areas mark the occurrence of prominences. 
 Black solid lines represent the light-curves determined from H$\alpha$ measurements whereas the blue solid lines represent the light-curves obtained from H$\beta$ measurements.}
\label{alphabetalightcurves}
 \end{center}
 \end{figure}

We obtained light-curves by integrating the spectral line profiles using fixed wavelength windows of 6555 to 6567\AA{} for H$\alpha$ and 4859.1 to 4865\AA{} for H$\beta$ and subtracting the continuum to obtain the line net fluxes. Light-curves can give insights into the status of the activity of the star. Especially on dMe stars, the H$\alpha$ line is dominated by chromospheric activity. Moreover, when looking for prominence oscillations the presence of a flare could be an indicator for a probable prominence eruption triggered by a flare. In Fig.~\ref{alphabetalightcurves} we show the light-curves of the nights when prominences have occurred on HK~Aqr. The light-curves have been normalized with the average H$\alpha$ or H$\beta$ flux, respectively, of the corresponding night. The grey shaded areas indicate the times when prominences are seen in absorption in Fig.~\ref{dynspecred}.
 In the second panel of Fig.~\ref{alphabetalightcurves} we show the night of the 08/08/2012 including prominences HK\_08\_P0 and HK\_08\_P2. Prominence HK\_08\_P0 lasts from spectrum no.~4-7, but also the spectra no.~1-2 indicate absorptions. Taking a look at Fig.~\ref{dynspecred} shows that there is indeed a weak absorption event visible, but it seems to be a separate prominence appearing before prominence HK\_08\_P0. Moreover, the quality of the spectra of this night is not as good as in the other nights, therefore a distinct identification becomes difficult. Spectrum no.~9 shows a sudden increase in flux which is related to a very low S/N.\\
The night of the 11$^{th}$ of August shows three prominences, namely HK\_11\_P1, HK\_11\_P2, and HK\_11\_P4, all of them are consistent with dips in the light-curve, the most pronounced event is HK\_11\_P2. The H$\beta$ light-curve of the same night shows several dips, the most pronounced event is also prominence HK\_11\_P2.\\
In the following night (12/08/2012) three prominences are visible (HK\_12\_P0, HK\_12\_P1, and HK\_12\_P2) which coincide with the dips in the light-curves. Spectrum no.~2 of that light-curve shows a peak which is related to the fact that guiding was lost at that time. As one can see from the light-curve, one would assign also spectra no.~3 to prominence HK\_12\_P0 but we do not see absorptions as signatures of prominences in the corresponding spectrum. From Fig.~\ref{allspec12} (see Appendix) on can see that spectrum no.~3 is very noisy. \\
As one can see, no typical flare light-curve is seen in Fig.~\ref{alphabetalightcurves}, except the event which was detected during the first night (see upper panel of Fig.~\ref{alphabetalightcurves}). Even there are enhancements in the light-curves of HK~Aqr in the night of the 11th of August (spectra no.~0-6 and spectra no.~29-33) and in the night of the 12th of August (spectra no.~28-30), we do not detect activity indicators such as the HeI line at 5876\AA{} \citep[e.g. ][]{Montes1997}, during those enhancements, which typically rises during flares.

\subsection{Prominence parameters}
\subsubsection{Prominence height}
\label{Resprom}
\begin{figure}
\begin{center}
\vspace*{0cm}
\includegraphics[width=7.0cm]{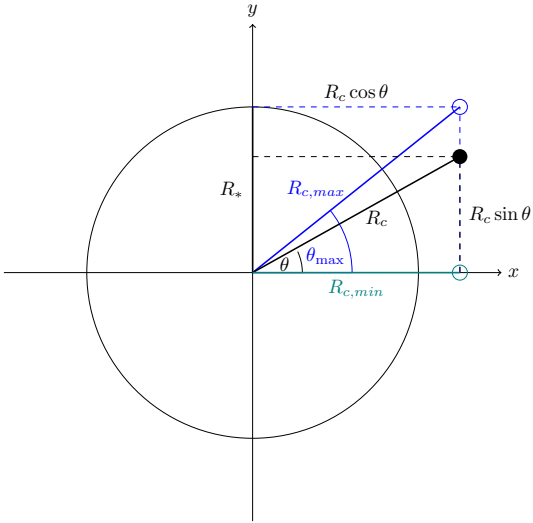}
 \caption{Illustration of the determination of $R_{cmax}$ and $R_{cmin}$ depending on $\theta_{min}$ and $\theta_{max}$. X-axis is in the line of sight of the observer and the y-axis marks the rotation axis of the star. No prominence showing absorption signatures can be detected above $\theta_{max}$, because it is then not projected onto the stellar disk.}
\label{geometrysketch}
 \end{center}
 \end{figure}

As in previous studies we assume that the prominence clouds are located at a distance $R_{c}$ from the stars center and rotate rigidly with the star. For simplicity we assume an inclination of $i=90^{\circ}$ which is likely considering the parameters of both stars \citep[see][]{Eibe1998, Maire2015}. For HK~Aqr we adopt a radius of 0.59R$_{\odot}$ \citep{Barnes2001} and for PZ~Tel a radius of 1.23R$_{\odot}$ \citep{Jenkins2012}. The cloud motion is then given by $V=\omega R_c\cos{\theta}$ where $\omega=2\pi/P$ is the stellar rotation rate and $\theta$ is the latitude of the cloud (see Fig.\ref{geometrysketch}). The product $R_c\cos{\theta}$ gives the distance of the cloud from the rotational axis. In the spectra we observe only the line-of-sight component of $V$ which is given by $V_c=V\sin{\phi}$ where $\phi=\omega(t-t_0)$ is the longitude, which is zero at $t=t_0$ when the cloud crosses the observer's meridian. Hence, the cloud's velocity evolution, as determined from the absorption features crossing the H$\alpha$ line, is described by \citep{CollierCameron1989a}
\begin{equation}
V_{c}=\frac{R_{c}}{R_{\star}} v \sin(i) \cos(\theta) \sin(\phi), 
\label{eq1}
\end{equation}
where we assume that $v\sin(i) \approx v_{eq} = \omega R_{\star}$. If t$\sim$t$_{0}$, then $\sin(\phi)\sim \phi$ and Eq.~\ref{eq1} describes the linear part of the radial velocity curve. To obtain the projected height $R_c\cos{\theta}$ we fit the linear part of the prominence tracks. For large-height prominences the prominence track lies completely in this linear part. Results are given in Table~\ref{Tab1} and shown in Fig.~\ref{osco} (see also Fig.~\ref{osco1} and Fig.~\ref{osco2} in the Appendix).\\
In Eq.~\ref{eq1} there are two unknown variables, which is the true height of the prominence $R_{c}$ and its latitude $\theta$, but only the combined quantity $R_c\cos{\theta}$ can be measured. However, one can deduce a range of possible $R_c$. Apparently, the minimum value of $R_c$ occurs if the clouds orbit in the equatorial plane where $\theta=0^{\circ}$ and is thus equal to the measured value $R_c\cos{\theta}$. The maximum of $R_c$ can be deduced from visibility arguments. For any true height $R_c$ there is a maximum possible latitude $\theta_{max}$ which the prominence can have and can still be seen in projection onto the stellar disk. Since all prominences we observe are seen in absorption it is obvious that their latitudes must be smaller than $\theta_{max}$. This maximum latitude of the clouds can be derived by the requirement that their heights from the equatorial plane must be smaller than $R_{\star}$, i.e. $\sin{\theta} \leq R_{\star}/R_{c}$ or $\tan{\theta} \leq R_{\star}/(R_{c}\cos{\theta})$. From the latter relation one can obtain the maximum $\theta$ for which a given prominence can still be seen against the stellar disk. Thus one can calculate the maximum possible height as $R_{cmax} = (R_c\cos{\theta})/\cos{\theta_{max}}$ (see Fig.~\ref{geometrysketch}). Since we obtained minimum values of $R_{c,min}<R_{\star}$ for several prominences it is possible to also estimate a minimum latitude for these cases through the requirement that $R_c \geq R_{\star}$ and thus $\cos{\theta_{min}} \geq R_c\cos{\theta}/R_{\star}$. 
From the results given in Table~\ref{Tab1} one can see that the obtained range of true heights is rather narrow. We find two types of prominences, large ones with $R_c \gg R_{\star}$ and small ones with $R_c \approx R_{\star}$. We find only large prominences for PZ Tel, but both large and small ones for HK Aqr. Large prominences must have latitudes smaller than about $40^{\circ}$, whereas the small prominences are located at high latitudes between $60-70^{\circ}$. To see whether the large prominences are stable structures with respect to gravity we calculate the co-rotation radii. For HK~Aqr we find a co-rotation radius of 3.16~R$_{\star}$ and for PZ~Tel a co-rotation radius of 3.42~R$_{\star}$. As one can see, except prominence HK\_08\_P0,
\noindent all other prominences are well below the co-rotation radius and are therefore expected to be stable. This finding is in agreement with \citet{Byrne1996} who found prominences with heights ranging from 0.34-2.4~R$_{\star}$ above the stellar surface. For the calculation of the co-rotation radius we adopted a mass of 0.4M$_{\odot}$ for HK~Aqr \citep{Byrne1996} and a mass of 1.13M$_{\odot}$ for PZ~Tel \citep{Jenkins2012}.\\

\subsubsection{Prominence area}
\label{promarea}
If the prominence material is optically thick we may estimate the projected area of the prominences. We follow the method in \citet{Dunstone2006b} to see if the prominences on HK~Aqr are optical thick or thin. We compare the H$\beta$/H$\alpha$ ratio of the prominence residuals to the theoretical curve of growth calculated in \citet{Dunstone2006b} for the Balmer lines. We see that our calculated values lie on the saturated part of the theoretical curve and are therefore assumed to be optical thick.
 \begin{figure}
\begin{center}
\vspace*{0cm}
\includegraphics[width=8.0cm]{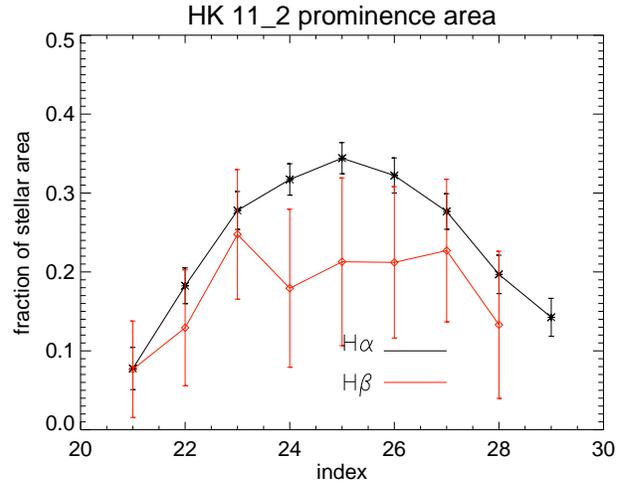}
 \caption{
 Prominence area derived for H$\alpha$ (black solid line) and H$\beta$ (red solid line). The area shows a rise when it is crossing the line of sight followed by a decay in area as the prominence moves off the stellar disk. The H$\beta$ area evolution follows the H$\alpha$ evolution only partly but the error bars are larger. }
\label{reslightcurve}
 \end{center}
 \end{figure} 
Now we can estimate the projected area. 
The equation which describes the flux density determined from the spectra closest to the line of sight crossing \citep[based on ][]{CollierCameron1990} is  
\begin{equation}
f_{\lambda}=f_{backgr, \lambda}(1-   \frac {A_{C}} {A_{star}}   [1-\exp(-\tau_{\lambda})]        ).
\label{Petragleichung}
\end{equation}
If the prominences are optical thick then one can neglect $\exp(-\tau_\lambda)$ and Eq.~\ref{Petragleichung} gives
\begin{equation}
\frac{A_{C}}{A_{\star}}=1-\frac{f_{\lambda}}{f_{backgr, \lambda}}.
\end{equation}
According to this equation we can infer the projected area of the prominences. In Table~\ref{Tab1} we list the derived projected areas. The values range from $\sim$7 ... 24\% for HK~Aqr and $\sim$ 4 ... 8\% for PZ~Tel. From Fig.~\ref{reslightcurve} one can see that the area derived from H$\alpha$ and H$\beta$ data are roughly consistent, but the errors for the area derived from H$\beta$ data are large, with respect to the errors for the area derived from H$\alpha$ data. 

\subsection{Indications of prominence oscillations}
\label{promosc}
From the variation of the center positions of the prominence tracks we detect periodic variability in two prominences on HK~Aqr (HK\_11\_P2 and HK\_12\_P1). The oscillatory motion is already visible when looking at the left panels of Fig.~\ref{osco}. To remove the stellar rotation we fitted the prominence track with a line and subtracted it (see section~\ref{stability} for details). Interestingly, we detect residual sinusoidal motion on HK~Aqr only, and none on PZ~Tel. In Fig.~\ref{osco1} and Fig.~\ref{osco2} (Appendix) we show sinusoidal motion of the prominence tracks of prominences 
HK\_11\_2 (in H$\beta$) and HK\_12\_1 (in H$\alpha$) which are the same prominence seen in different nights. 
We also searched for variability of the center-positions in intensity but none could be found.\\  
The prominence tracks showing oscillatory motion have been fitted with a sinusoidal function 
of the form
\begin{equation}
d(t) = x_{0}\sin \bigg(\frac{2 \pi t}{P} + \phi \bigg)+d_{0}, 
\label{Vrsnakglgsin}
\end{equation}
where x$_{0}$ is the amplitude, P the period, $\phi$ the phase, and d$_{0}$ a displacement in the y-axis. As the tracks of prominences HK\_11\_2, and HK\_12\_1 consist of 8 and 6 data points, respectively, we do not include damping, as the number of free parameters of a damped sine function is close to the number of data points of prominences HK\_11\_2, and HK\_12\_1.
To exclude that the detected variations might come from instrumental effects we perform the following test. Photospheric lines are assumed to stay stable during the time series with respect to their line center. We choose the CaI line at 6439\AA{} which we fitted in the same way as the residual spectra. The test showed that the center positions show deviations within one wavelength resolution element. As the spectra are oversampled this corresponds to $\sim$2~km~s$^{-1}$. Therefore we exclude the possibility that the periodic variability of the center-positions are influenced 
\renewcommand{\thefootnote}{\alph{footnote}}
\begin{table*}
\centering
 \begin{minipage}{180mm}
   \caption{Parameters of prominences detected on HK~Aqr and PZ~Tel during the 07-12$^{th}$ of August 2012. For each detected prominence we list the phase, the event type (if we see the full prominence transit or only a part of it), the R$_{c}$cos($\theta$) value which corresponds also to R$_{c\_min}$ (for $\theta$=0$^{\circ}$), the maximum height (as R$_{c}$ and R$_{c\_max}$ are counted from the center of the star one has to subtract the stellar radius to obtain the height of the prominences above the stellar surface), and the projected area. For the detected periodic variability of center-positions we list amplitude and period. We list amplitudes and periods for solutions (a) and (c). For information see text. For some of the partially observed events we could not derive their height as there were too few data points to be fitted.}
   \label{Tab1}
  \begin{tabular}{cccccccccc}
  \hline
struct    &    date     &    phase    &  event type & R$_{c}$cos($\theta$) &  R$_{c\_max}$ &   area      & oscillation &           amplitude    &          period         \\
          &             &             &             &   [R$_{\star}$]      & [R$_{\star}$] &    [\%]     &             &         [km~s$^{-1}]$  &           [min]         \\
      \hline
HK 08 P0  & 08-08-2012  &  0.45-0.60  &     full    &        3.75          &      3.88     &    13.3     &     no      &               -        &             -            \\
HK 08 P2  & 08-08-2012  &  0.64-0.74  &    partial  &        0.32          &      1.05      &    19.7     &     no      &               -        &             -           \\
      \hline
HK 09 P1  & 09-08-2012  &  0.17-0.22  &    partial  &           -          &       -       &      -      &     no      &              -         &             -            \\
      \hline
HK 11 P1  & 11-08-2012  &  0.40-0.62  &     full    &        0.49          &      1.11     &     9.4     &      no     &             -          &           -              \\
HK 11 P2  & 11-08-2012  &  0.64-0.74  &     full    &        1.67          &      1.95     &    24.1     &     yes     &   6.26$\pm$0.92(a)    &    50.83$\pm$3.02(a)    \\
          &             &             &             &                      &               &             &             &   6.59$\pm$1.62(c)    &    61.21$\pm$7.52(c)    \\
HK 11 P4  & 11-08-2012  &  0.87-0.93  &    partial  &           -          &      -        &      -      &      no     &             -          &           -             \\
      \hline
HK 12 P0  & 12-08-2012  &  0.53-0.62  &     full    &        2.41          &     2.61      &     7.4     &      no     &             -          &           -             \\
HK 12 P1  & 12-08-2012  &  0.66-0.78  &     full    &        0.54          &     1.14      &    16.8     &     yes     &    5.44$\pm$1.40(a)   &    46.25$\pm$4.35(a)     \\
          &             &             &             &                      &               &             &             &    8.15$\pm$1.62(c)    &    45.63$\pm$3.45(c)    \\
HK 12 P2  & 12-08-2012  &  0.84-0.94  &     full    &        1.35          &     1.68      &    15.9     &      no     &             -          &           -              \\
          \hline
          \hline
struct    &    date     &    phase    & event type  & R$_{c}$cos($\theta$) &  R$_{c\_max}$ &   area      & oscillation &          amplitude     &           period         \\
          &             &             &             &    [R$_{\star}$]     & [R$_{\star}$] &   [\%]      &             &         km~s$^{-1}$    &           [min]          \\
PZ 07 P1  & 07-08-2012  &  0.9-0.17   &   partial   &       -              &       -       &     -       &     no      &              -         &              -           \\
      \hline
PZ 08 P1  & 08-08-2012  & 0.020-0.085 &    full     &     1.44             &    1.75       &    4.1      &     no      &              -         &              -           \\
      \hline
PZ 09 P1  & 09-08-2012  & 0.150-0.252 &    full     &     1.05             &    1.45       &    7.8     &     no      &              -         &               -          \\
      \hline
PZ 10 P0  & 10-08-2012  & 0.144-0.248 &    full     &     -                &       -       &     -       &     no      &              -         &               -          \\
      \hline
PZ 11 P0  & 10-08-2012  & 0.31-0.33   &   partial   &       -              &       -        &     -       &     no     &              -         &               -          \\
      \hline
PZ 12 P0  & 12-08-2012  & 0.299-0.328 &   partial   &        -             &       -        &     -       &     no     &              -         &              -           \\
      \hline
\end{tabular}
\end{minipage}
\end{table*} 

\noindent by instrumental effects. In Table~\ref{Tab1} we show the parameters of the sinusoidal motions of the H$\alpha$ prominence tracks.\\
 \begin{figure}
\begin{center}
\vspace*{0cm}
\includegraphics[width=8.0cm]{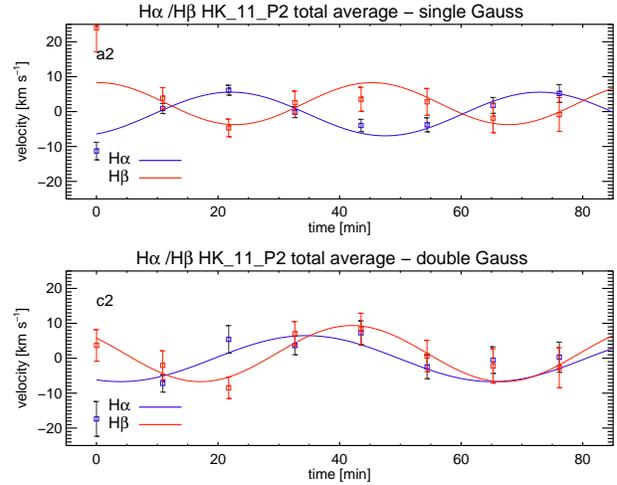}
 \caption{H$\alpha$ (blue square symbols) and H$\beta$ (red square symbols) prominence tracks for prominence HK\_11\_P2 determined from single gaussian fitting (upper panel) and double gaussian fitting (lower panel). Overplotted is a sinusoidal function (Eq.~\ref{Vrsnakglgsin}). For the prominence track derived from single gaussian fitting we clearly see an anti-phase behaviour of H$\alpha$ and H$\beta$. For the prominence track derived from double gaussian fitting we see a phase shift rather than an anti-phase behaviour.}
 \label{halphabetacomp}
 \end{center}
 \end{figure} 
The derived periods of periodic variability of prominences HK\_11\_P2 and HK\_12\_P1 lie within the range of solar large amplitude oscillations, whereas the derived amplitudes are lower and lie between solar small- and large-amplitude oscillations. The derived amplitudes are close to the spectral resolution of FEROS ($\sim$6~km~s$^{-1}$).\\
As we have detected prominences HK\_11\_P2 and HK\_12\_P1 also in H$\beta$ we are able to infer if both show also periodic variability in H$\beta$. By comparing the H$\alpha$ and H$\beta$ tracks for prominence HK\_11\_02  we see that H$\beta$ shows the H$\alpha$ oscillation in anti-phase (see Fig~\ref{halphabetacomp}), at least for prominence tracks derived from single gaussian fitting from residuals obtained from total average subtraction (we find the same results for residuals obtained from daily average subtraction). For prominence tracks derived from double gaussian fitting, for both residuals determined from daily and total average subtraction, we find a phase shift between H$\alpha$ and H$\beta$ rather than an anti-phase behaviour.
For the HK\_12\_01 event we can not reproduce the sinusoidal motion in H$\beta$, as seen in H$\alpha$. The H$\alpha$ and H$\beta$ discrepancy seen in HK\_11\_P2 is possibly related to the different quality of the data around H$\alpha$ and H$\beta$ or is a sign of a prominence oscillation seen in different regions in the same prominence (see section~\ref{oscdissc}).

\subsubsection{Stability of the periodic variability, errors, and significance}
\label{stability}
\begin{figure}
\begin{center}
\vspace*{3cm}
\includegraphics[width=17.0cm]{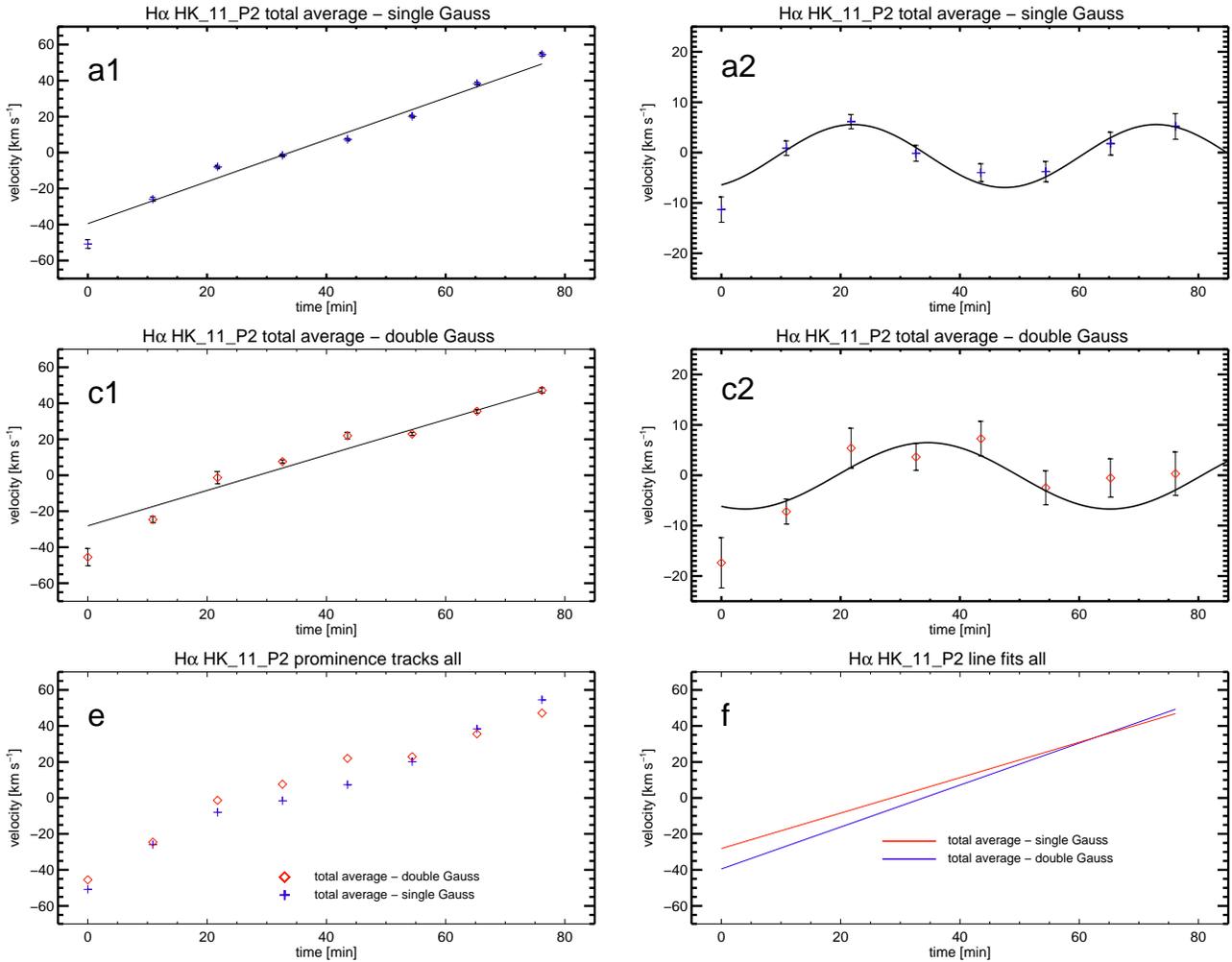}
 \caption{Panels in the left column from top to bottom: Prominence H$\alpha$ tracks of prominence HK\_11\_P2 derived from residuals generated from (a1) total average subtraction and single Gauss fitting and (c1) total average subtraction and double Gauss fitting. In panels a1 and c1 the linefits representing the cloud velocity are over plotted as black solid lines. Panel (e) shows an overplot of prominence tracks a1 and c1. Panels in the right column from top to bottom: Prominence H$\alpha$ tracks of prominence HK\_11\_P2 corrected for cloud velocity derived from residuals generated in the same order as for the panels (a1-c1) in the left row. Overplotted is a sinusoidal function from which we determine period and amplitude of the periodic variability. In the last panel in the right column (f) we show the spread of the line fits (a1-c	1) used for removing the cloud velocity.}
\label{osco}
 \end{center}
 \end{figure}

As we deduce sinusoidal motion of the center positions of the residuals, the generation of the residuals might have an influence on the parameters of periodic variability of the center positions of the prominence tracks. 
To check what the influence of different built residuals is, we build total and daily averages. To check what the influence of different fitting functions is we use single and double gaussian functions to fit the residuals. 
This results in four cases, (a) single gaussian fitting of the residual built from total average subtraction, (b) single gaussian fitting of the residual built from daily average subtraction, (c) double gaussian fitting of the residual built from total average subtraction, and (d) double gaussian fitting of the residual built from daily average subtraction. The influence of daily and total averaging is smaller with respect to variations in the center positions compared to the influence of single and double gaussian fitting, respectively. Therefore we plot in Fig.~\ref{osco} the H$\alpha$ prominence tracks determined for cases (a) and (c) for prominence HK\_11\_P2 in the left panels. The prominence tracks are over-plotted with the line fits from which we determine the prominence height and which is subtracted to remove the cloud velocity. The prominence tracks with removed cloud velocity are shown in the panels in the right column of Fig~\ref{osco}. Over-plotted is the sinusoidal function (see Eq.~\ref{Vrsnakglgsin}). One can see that case (a), which originates from single gaussian fitting, shows a sinusoidal motion. Case (c), which originates from double gaussian fitting, can be fitted with a sine function, showing  phase shift and a difference in amplitude, with respect to case (a).\\
Going back to Fig.~\ref{showspec} one can see that single and double gaussian fitting applied to residuals derived from daily and total average subtraction yield different center positions. Both fitting functions introduce systematic errors, as some residuals are better fitted by single gaussians and some are better fitted by double gaussians. In the analysis we introduced a weighted mean to include all four cases ((a)-(d)), but this representation includes also all systematic errors, therefore we believe that a representation of the prominence tracks by a weighted mean is not appropriate.\\
To evaluate the goodness of fit of the sine function we calculate its reduced $\chi^{2}$. In addition we fit a line to the data and calculate its reduced $\chi^{2}$ for comparison.
Especially as the amplitudes are rather small and the data errors are not negligible the choice of a straight line as comparison fit function is reasonable. In table~\ref{significance} we list the reduced $\chi^{2}$ values for prominences HK\_11\_P2 and HK\_12\_P1 for residual cases (a) and (c). The reduced $\chi^{2}$ depends on the number of observations (i.e. data points) and free parameters of the fitting function. For prominence HK\_11\_P2 there are only 8 data points and for prominence HK\_12\_P1 only 6. Therefore the usage of a damped sine function is not applicable especially for prominence HK\_12\_P1, which has 6 data points only. \\
The different reduced $\chi^{2}$ values of prominence HK\_12\_P1 clearly show that fitting with a sine function is reasonable as the reduced $\chi^{2}$ values are in all cases smaller than for line fitting.\\
The different reduced $\chi^{2}$ values of prominence HK\_11\_P2 (both H$\alpha$ and H$\beta$) show that fitting with a sine function is reasonable. 
All reduced $\chi^{2}$ values for prominence HK\_11\_P2 show that the sine function matches the observed values better than a line. 
The differences of deriving center positions using single and double Gauss fitting can be inferred from Fig.~\ref{showspec} where we show the residual spectral sequences of prominences HK\_11\_P2 and HK\_12\_P1 over-plotted with single- and double-gaussian fits. The double Gauss functions follows the absorptions in the residuals in more detail than the single Gaussian functions. 
If we take a look at the large discrepancies in position of prominence HK\_11\_P2, which are residual no.~24 and no.~25, then one recognizes that in the blue wing of the H$\alpha$ residual an absorption has formed, which turns into emission in the following spectra.  
\begin{table}
\centering
   \caption{Reduced $\chi^2$ calculated for prominences HK\_11\_P2 (H$\alpha$ and H$\beta$) and HK\_12\_P1 (H$\alpha$). 
   In each column we list first the reduced $\chi^2$ value derived from fitting with a sine function and separated by a slash from fitting with a line. 
   }
   \label{significance}
\begin{tabular}{cccc}
                   &  HK\_11\_P2 H$\alpha$  &  HK\_11\_P2 H$\beta$   &  HK\_12\_P1 H$\alpha$     \\
\hline  
                   &     sine/line          &     sine/line           &      sine/line             \\ 
 $\chi^2_{a}$      &   2.21/10.4            &    16.16/18.02          &      2.86/6.26             \\   
 $\chi^2_{c}$      &     3.99/4.45          &    1.38/24.30           &      9.36/10.18            \\
                   &                        &                   \\
\hline

\end{tabular}
\end{table}
\subsection{Estimation of magnetic field strength in stellar prominences: stellar coronal seismology}
\label{magnet}
Observed oscillations can be used to estimate plasma parameters in the prominences. The periodic variability of the center-positions is only found in line-of-sight velocity but not in intensity, which indicates their transverse character. Then the periodic variability of the center positions of the prominences can be interpreted in terms of Alfv\'en or magneto-hydrodynamic kink waves. Their dispersion relation depends on magnetic field strength and plasma density. The wavelength can be assumed from the spatial extent of prominences. Then the magnetic field strength can be estimated if one assumes a plasma density in the prominences. \citet{Dunstone2006b} used the curve of growth method to obtain the column density of H$\alpha$ and CaII atoms. Both are necessary to estimate the column density of hydrogen atoms in the ground state. By assuming that at prominence temperatures of 10$^{4}$~K the Ca atoms are single ionized, these authors used the hydrogen to calcium ratio for solar prominences. By doing so they obtain the column density for hydrogen atoms in the ground state. The data \citet{Dunstone2006b} used have a much higher S/N as our data and therefore we can not repeat their method to obtain the column density of hydrogen in the ground state. But as an approximation we use our determined ratio of EW(H$\beta$)/EW(H$\alpha$) (0.37) and determine the logarithmic hydrogen column density for state N$_{2}$ from the curve of growth given in \citet{Dunstone2006b} which gives logN$_{2}$ $\sim$ 18~m$^{-2}$. As we have no CaII~H$\&$K measurements we adopt the N$_{1}$ to N$_{2}$ ratio of Speedy Mic, which we assume to be comparable to the ratio of HK~Aqr and find a log(N$_{1}$) value of 23.7~m$^{-2}$. As we have estimated the area of the prominences we are able to calculate the mass according to M=m$_{H}$N$_{1}$A. For the only prominence on HK~Aqr for which we have H$\alpha$ and H$\beta$ measurements and which shows sinusoidal motion in both lines, which is HK\_11\_P2, we calculate a mass of 5.7$\times$10$^{16}$g. To obtain the density of the prominence plasma we have to assume the extension in the third dimension of the prominence. We assume  that the thickness is comparable to the length scale which we estimate as $\sqrt{A_c/A_\star}$.
Applying this geometrical thickness to the prominence area gives a volume of $\sim$2$\times$10$^{31}$cm$^{3}$. The hydrogen density of the prominence plasma results finally to 2.9$\times$10$^{-15}$g~cm$^{-3}$. In Table~\ref{Tab2} we list derived parameters for HK\_11\_P2 and for a solar prominence analysed in \citet{Schwartz2004}, as comparison. One can see that the solar and stellar values differ within one or two orders of magnitude. Of course, the scaling of the prominence thickness is the most uncertain parameter and influences the results significantly.\\
\begin{table}
\centering
   \caption{Comparison of solar and stellar prominence parameters. The stellar prominence parameters are estimated from HK\_11\_P2 and the solar parameters are taken from \citet{Schwartz2004}.}
   \label{Tab2}
  \begin{tabular}{cccccc}
\hline
                              &         HK~Aqr          &       Sun              \\
                              &                         &                        \\ 
mass density [g~cm$^{-3}$]    &  2.9$\times$10$^{-15}$  &    10$^{-15}$          \\
mass  [g]                     &  5.7$\times$10$^{16}$   &    10$^{14}$-10$^{15}$ \\
volume [cm$^{3}$]             &  2.0$\times$10$^{31}$   &    $\sim$10$^{29}$     \\
area [cm$^2$]                 &  7.4$\times$10$^{20}$   &   9.9$\times$10$^{19}$ \\
thickness [cm]                &  2.7$\times$10$^{10}$   &    7$\times$10$^{9}$   \\
\hline
\end{tabular}
\end{table} 

\noindent To assess the magnetic field of prominence HK\_11\_P2 we assume kink oscillations of a cylindrical fluxtube with length L. We assume that the density is much higher in prominences than in the surrounding coronal plasma as it is the case on the Sun. Then the phase speed of the kink oscillation is defined \citep{Edwin1983} as
\begin{equation}
c_{k} = \frac{\sqrt{2} B_{0}}{\sqrt{4\pi\rho_{0}}} = \frac{2L}{T}.
\label{kinkspeed}
\end{equation}
B$_{0}$ is the magnetic field of the prominence plasma, $\rho_{0}$ the mass density of the prominence plasma, L denotes the length of the prominence and T is the period of the first harmonic of the prominence oscillation. For the calculation of the magnetic field strength of HK\_11\_P2 we take the values given in Table~\ref{Tab2}. The only missing parameter is the length of the prominence, which we have deduced assuming that the prominences area has a quadratic shape. 
The average area of prominence HK\_11\_P2 is  7.4$\times$10$^{20}$ cm$^{2}$ and the length of a corresponding square with the same area is 2.7$\times$10$^{10}$~cm. 
Using these values we get a B$_{0}$ of $\sim$2~G. Since the prominence length is underestimated because prominences are known to be elongated, B$_{0}$ is a lower limit. Prominence HK\_11\_P2 is located at almost 1~R$_{\star}$ from the stellar surface. This means that the magnetic field strength is  estimated approximately at this height, which is quite high in comparison with solar prominence heights. Recently, \citet{Zaqarashvili2013} estimated solar coronal magnetic field strengths through coronal seismology using radio observations and found $\sim$ 1~G at the height of 1 R$_{\sun}$. Therefore, our estimations are in good coincidence with those of the solar corona. This similarity is surprising as surface magnetic fields of dM stars are much higher than on the Sun \citep{Mullan1975}. 
\section{Discussion}
\label{discussion}
In this section we will discuss the periodic variability of the center-positions of two prominences detected on HK~Aqr, especially the anti-phase detection in H$\beta$ of prominence HK\_11\_P2. Moreover we will discuss the roughly estimated magnetic field strength of the prominence.
\subsection{Indications of prominence oscillations}
\label{oscdissc}
On the Sun one distinguishes between small amplitude oscillations which are mainly related to quiescent prominences with amplitudes of a few km~s$^{-1}$ and large amplitude oscillations mainly related to flares with amplitudes of $>$ 20~km~s$^{-1}$.\\
On stars there is so far no detection of prominence oscillations of this kind found here. \citet{Houdebine1993} found during a flare on the dMe star AD~Leo that the CaII~K line revealed periodic Doppler shifts (of the whole line) with a period of a few minutes and a maximum amplitude of 53~km~s$^{-1}$. \\
\citet{Jeffries1993} analyse Speedy~Mic spectra and find prominences co-rotating with the star. Looking at their Fig.~3 one can see variations of the absorption transients around the radial velocity fit, which the authors mention as ``... there is marginal evidence for a sinusoidal velocity for transient B.'' Probably we see here the first manifestation of prominence oscillations on Speedy~Mic.\\
 With respect to spectrographs with moderately high resolving power, such as FEROS, one will be able to resolve the stellar analogue of solar large amplitude oscillations only. As we detected amplitudes with velocities lying between solar small- and large amplitude oscillations on HK~Aqr a higher resolving power would have been desirable.\\ 
As mentioned above, if we indeed detected the stellar analogue of the solar large amplitude oscillations then one would expect a flare and/or CME occurring as the exciter for the oscillation, as the oscillation of huge masses requires much energy \citep[e.g.][]{Liu2013}. No optical signature of a flare and/or a CME was detected before prominence HK\_11\_P2.\\
As mentioned in section~\ref{promosc} prominence HK\_11\_P2 shows an anti-phase behaviour of periodic variability of the center-positions in H$\beta$. From the upper panel of   Fig.~\ref{halphabetacomp}, which

\onecolumn
  \twocolumn

\noindent shows the H$\alpha$ and H$\beta$ prominence tracks derived from single gaussian fitting, one can see that the H$\beta$ oscillation is indeed in anti-phase to the H$\alpha$ oscillation. The H$\alpha$ and H$\beta$ prominence tracks derived from double gaussian fitting (lower panel of Fig.~\ref{halphabetacomp}) show a reduced phase shift. We already introduced the problem of systematic errors in section~\ref{stability} for single and double gaussian fitting. Probably this might become an issue also here. However, in the following we discuss this finding with a special emphasis to solar observations of prominence oscillations.\\
The neutral prominence matter is mainly comprised of hydrogen atoms of different excitation levels. H$\alpha$ and H$\beta$ are generated from the same atomic level (n=2) and from similar temperatures. If we see in H$\alpha$ and H$\beta$ different regions of the prominence then it could be possible that we see oscillations out of phase. On the Sun it is possible to spatially resolve filaments and make the fine structure, the so-called fibrils, visible \citep{Schmieder1989}. Regarding small-amplitude oscillations it is known that individual fibrils, or even groups of fibrils of the filament oscillate independently \citep{Diaz2003}. \citet{Terradas2002} found anti-phase oscillations in a sub-region of a quiescent prominence, detected in H$\beta$ filtergrams with a period of 75~minutes and a mean amplitude of 2~km~s$^{-1}$. The anti-phase was found in different regions of the prominence but not in different wavelengths, as for HK~Aqr. \citet{Ning2009} report on a solar filament, observed with H$\alpha$ and CaII filters, consisting of two spines which oscillated in anti-phase. The oscillation amplitude was measured to be on average 3~km~s$^{-1}$ which is in the range of small amplitude oscillations but the period of 98~minutes classifies it as a long-period oscillation. 
\citet{Luna2008} numerically investigated transverse oscillations in two parallel cylindric slabs to understand the collective oscillatory behaviour of the system. Different modes were found with oscillations in and out of phase. With this model several solar findings with respect to prominence oscillations can be explained.\\ 
\citet{Zapior2015} simultaneously observed solar quiescent prominences in several spectral lines, including H$\alpha$ and H$\beta$. These authors detected small amplitude oscillations of quiescent prominences showing the same phase in H$\alpha$ and H$\beta$. Again, there are no simultaneous observations of H$\alpha$ and H$\beta$ solar prominence large amplitude oscillations. 
However, stellar prominences are, compared to solar ones, huge structures. They are detectable in several spectral lines including H$\alpha$ and H$\beta$. It is likely that the fine-structure of stellar prominences is also comprised of solar analogues of fibrils. Possibly those are also able to oscillate out of phase.  Furthermore, it could be possible that not only the whole loop containing the prominence oscillates, but also the prominence H$\alpha$ and H$\beta$ plasma itself, producing a superimposed motion. \\
The H$\beta$ anti-phase oscillation was detected in one event only and only for case (a), case (c) showed a reduced phase shift only. Additional observations with higher spectral resolution of the H$\alpha$ and H$\beta$ oscillatory signatures are needed to draw conclusions on this behaviour.\\
On PZ~Tel we did not detect sinusoidal motion in the prominence tracks, only non-periodic variability. Possibly PZ~Tel has oscillations with amplitudes not detectable with our spectral resolution.\\
In section~\ref{stability} we have shown the influence of different built residuals and different fitting functions on the variation of the center-positions. One source of uncertainty for the variation of the center-positions of the prominences is the finding of a reference profile, which is subtracted to receive the residual spectra from which the center-positions are determined. Ideally one would need a spectrum which is not influenced by any prominence signature and/or signature of stellar activity at the same time. From Fig.~\ref{dynspecred} one can see that for HK~Aqr no ``quiescent'' spectrum is available, either it is affected by absorptions and/or emission features in the residuals.\\
Young main-sequence stars have high activity levels and frequent flaring is often observed, especially for dMe stars. A source of uncertainty which possibly influences the center-positions of the residuals is the activity of the star. In the first night of the observations we have detected a flare on HK~Aqr which is seen as distinct enhancement of the H$\alpha$ profile. Such an enhancement has not been detected again in the remaining nights (see Fig.~\ref{alphabetalightcurves}). A flare event leads to an increase of the flux of an H$\alpha$ profile, even if a prominence is projected onto the disk a pronounced flare event (like the one seen during the first night) can be still identified from the corresponding light-curve, therefore we are confident that the only pronounced flare detected in the data is that from the first night. The identification of less pronounced flares is problematic as they cause changes in the light-curves at a flux level comparable to prominences or even weaker. If such less pronounced flares have occurred and are superimposed on the prominence signatures then their contribution is hard to estimate, because the amount of residual flux varies with longitude and latitude of prominence occurrence. Moreover, if a plage region is also visible at the time of prominence occurrence then this is also seen superimposed onto the residual, as it is the case for prominence HK\_11\_P2 where we see an emission emerging in residual no.~27 in Fig.~\ref{allspec11}. We do not expect that a superimposed plage region causes periodic variability, but a deviation from the linear prominence track.\\
For the superposition of a flare it is nearly impossible to disentangle flare and prominence contribution. A flare causes the H$\alpha$ core to symmetrically rise and line asymmetries related to a flare are known to occur in the wings (both blue and red) of H$\alpha$. Prominences can be seen on HK~Aqr in absorption with a wavelength range of $\pm$1.5\AA{} from line center which corresponds to the $v$sin$i$ of HK~Aqr of $\sim$70~km~s$^{-1}$. From the Sun, it is known that asymmetries during flares related to up- and down-ward flows can reach several tens of km~s$^{-1}$ \citep{Ichimoto1984}. We can not determine the influence of frequent less energetic flares, compared to the event during the first night, on the determination of the center-positions derived from the residuals, as we can not identify them from the light-curves. The determination of underlying stellar activity would require spectroscopic data of HK~Aqr without prominence signatures or spectroscopic data of a stellar twin of HK~Aqr showing no prominences.\\

\subsection{Magnetic field of stellar prominences}
In section~\ref{magnet} we derived under several assumptions for the prominence geometry the magnetic field strength of prominence HK\_11\_P2. We derived a lower limit of B$_{0}\sim$2~G which is comparable to magnetic fields in solar prominences which are in the range of a few G \citep{Parenti2014} but are located at lower heights compared to the height of prominence HK\_11\_P2. Therefore the centrifugal forces acting on the prominence material is smaller. \citet{CollierCameron1989a} found prominences on AB~Dor extending beyond the co-rotation radius therefore they argued for an additional force (magnetic tension) acting against the centrifugal force. The prominences detected on HK~Aqr are below the co-rotation radius, which is in agreement with \citet{Byrne1996} and should be therefore stable. \\
\citet{Wang2010} performed an automated statistical analysis of solar prominences and found that the majority reaches a height of 2.6$\times$10$^{4}$km. Heights of stellar prominences are much bigger than solar ones. From literature (see section~\ref{Intro}) and the height determination from this study stellar prominences reach heights of a few stellar radii. E.g. for prominence HK\_11\_P2 we find a maximum height above the stellar surface of 0.67~R$_{\star}$ which corresponds to $\sim$2.8$\times$10$^{5}$~km which is a factor $\sim$11 higher than for average solar prominences. \citet{CollierCameron1989a} used an approximation to estimate the magnetic field strength inside the prominences on the fast rotating K dwarf AB~Dor. Saying that magnetic pressure must be bigger than the gas pressure,
\begin{equation}
\frac{B^{2}}{8\pi} > nkT
\label{Collglg}
\end{equation}
these authors estimated a magnetic field strength of the AB~Dor prominences of $\sim$1~G assuming a neutral particle density n of 10$^{9}$~cm$^{-3}$ which is in agreement with our estimate of particle density (3.9$\times$10$^{8}$~cm$^{-3}$) and magnetic field strength (as estimated from Eq.~\ref{kinkspeed}). If we calculate the magnetic field strength of the prominence using Eq.~\ref{Collglg} and assuming a temperature of 10$^{4}$~K we get a lower limit for the magnetic field strength of 0.12~G. We also consider the contribution from ions and electrons, assuming that both are equal in number, to obtain the final magnetic field strength. Therefore we multiply the obtained magnetic field strength by a factor of 3 assuming the same number densities of ions and neutral hydrogen atoms. This gives 0.36~G  which is similar to the B$_{0}$ derived for prominence HK\_11\_P2.\\
\citet{Vrsnak1984} investigated vertical prominence oscillations and argued that the magnetic field of the prominence must satisfy the relation of kinetic energy density and magnetic pressure 
\begin{equation}
\frac{B^{2}}{8\pi} > \frac{\rho v^{2}}{2}.
\label{Vrsnakglg1}
\end{equation}
Here $v$ is the amplitude of the oscillation. From this equation we calculate a lower limit of the magnetic field strength of 0.12~G. \citet{Houdebine1993} estimated for their CaII~K Doppler-shifts a minimum magnetic field strength of $\sim$20~G assuming an electron density of 5$\times$10$^{11}$cm$^{-3}$.\\
The above estimations have shown that the determined magnetic field strength in the prominence deduced from the oscillations agrees with magnetic field strengths estimated by \citet{CollierCameron1989a} for prominences on AB~Dor. Moreover the deduced magnetic field strength fulfils the requirements expressed in Eq.~\ref{Collglg} and \ref{Vrsnakglg1} and agrees with the solar coronal magnetic field strength at similar heights \citep{Zaqarashvili2013}.
\section{Conclusions}
We have analysed 6 nights of spectroscopic observations of the fast rotating, young, and active stars HK~Aqr and PZ~Tel. We detected in total 14 prominences on both stars. Two prominences detected on HK~Aqr show periodic variability of the center-positions of prominence tracks with periods in the range of solar large amplitude oscillations. The amplitudes are smaller than found for large-amplitude oscillations in solar prominences. In one of the prominences of HK~Aqr we detect an anti-phase oscillation in H$\beta$ which probably is a sign of fine structure of stellar prominences.\\
Using several assumptions we estimated the magnetic field strength of a prominence detected on HK~Aqr which is in the order of so far reported magnetic field strengths in stellar prominences. \\

\section*{Acknowledgements}

ML, PO, and AH acknowledge the support from the FWF project P22950-N16. AH acknowledges the support from the FWF project P27765-N27. PO and HL acknowledge the support by the FWF project P 27256-N27. HL acknowledges the support by the FWF NFN project S116 ``Pathways to Habitability: From Disks to Active Stars, Planets and Life'', and the related FWF NFN subproject S116607-N16. The work of TZ was supported by the Austrian “Fonds zur F\"{o}rderung der Wissenschaftlichen Forschung” under project P26181-N27 and by FP7-PEOPLE-2010-IRSES-269299 project- SOLSPANET.

\bibliographystyle{mn2e_update}
   \bibliography{Mybibfile}
   \appendix

\section{}
      \onecolumn

\begin{figure}
\begin{center}
\vspace*{3cm}
\includegraphics[width=17.0cm]{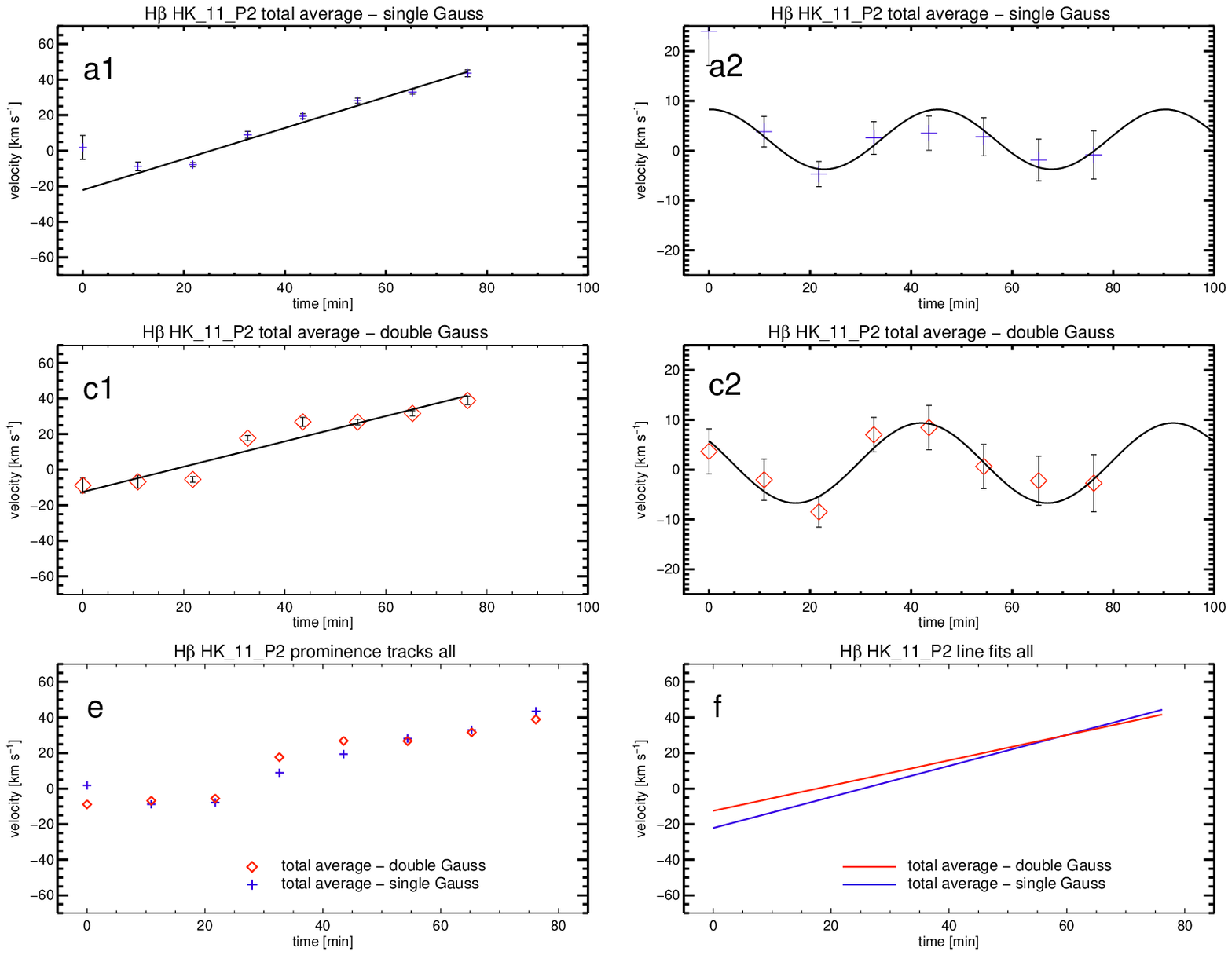}
 \caption{Same description as for Fig.~\ref{osco}, but for H$\beta$.}
\label{osco1}
 \end{center}
 \end{figure}
\onecolumn

\begin{figure}
\begin{center}
\vspace*{3cm}
\includegraphics[width=17.0cm]{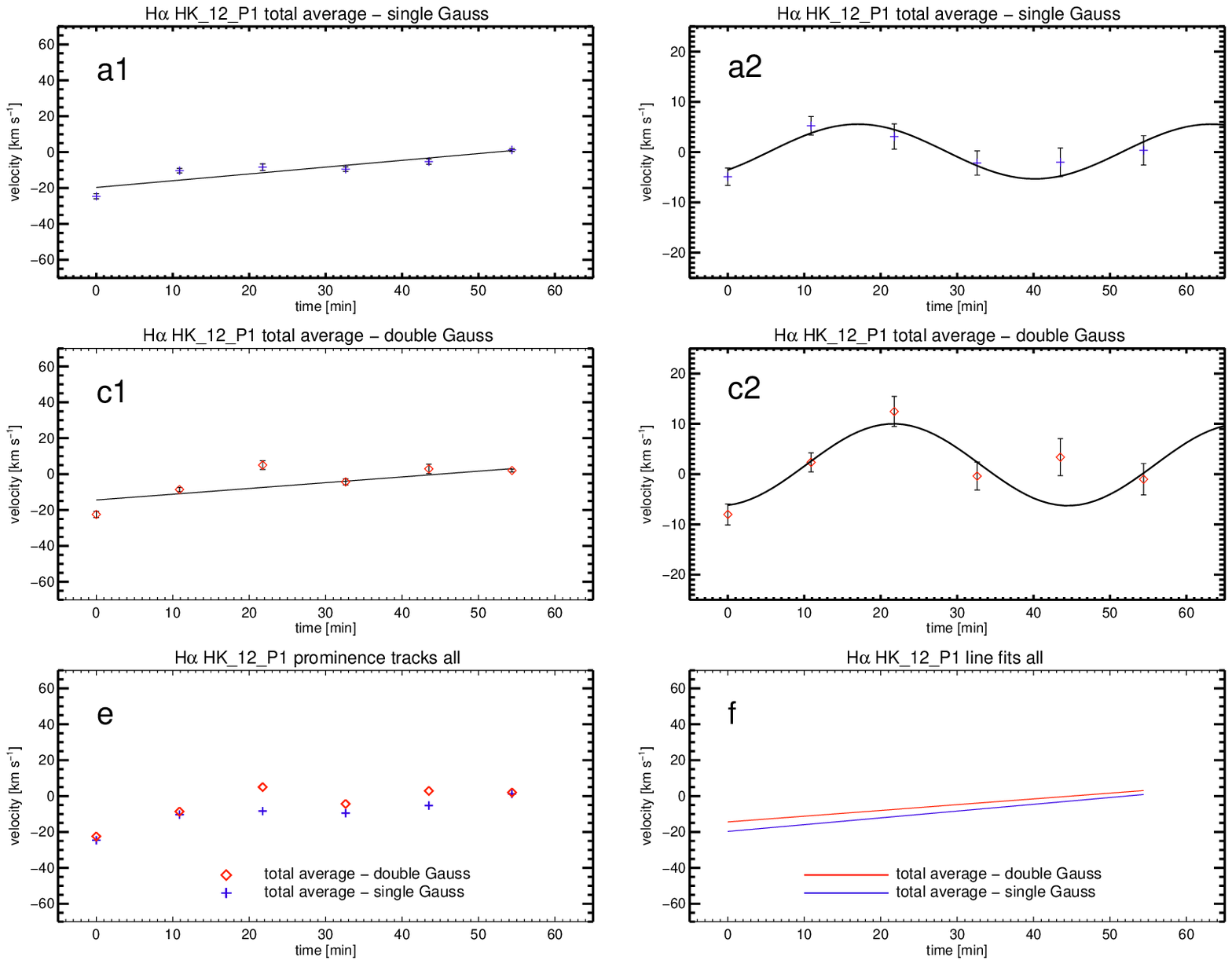}
 \caption{Same description as for Fig.~\ref{osco}, but for prominence HK\_12\_P1.}
\label{osco2}
 \end{center}
 \end{figure}

\onecolumn

\begin{figure}
\begin{center}
\vspace*{0cm}
\includegraphics[width=10.0cm]{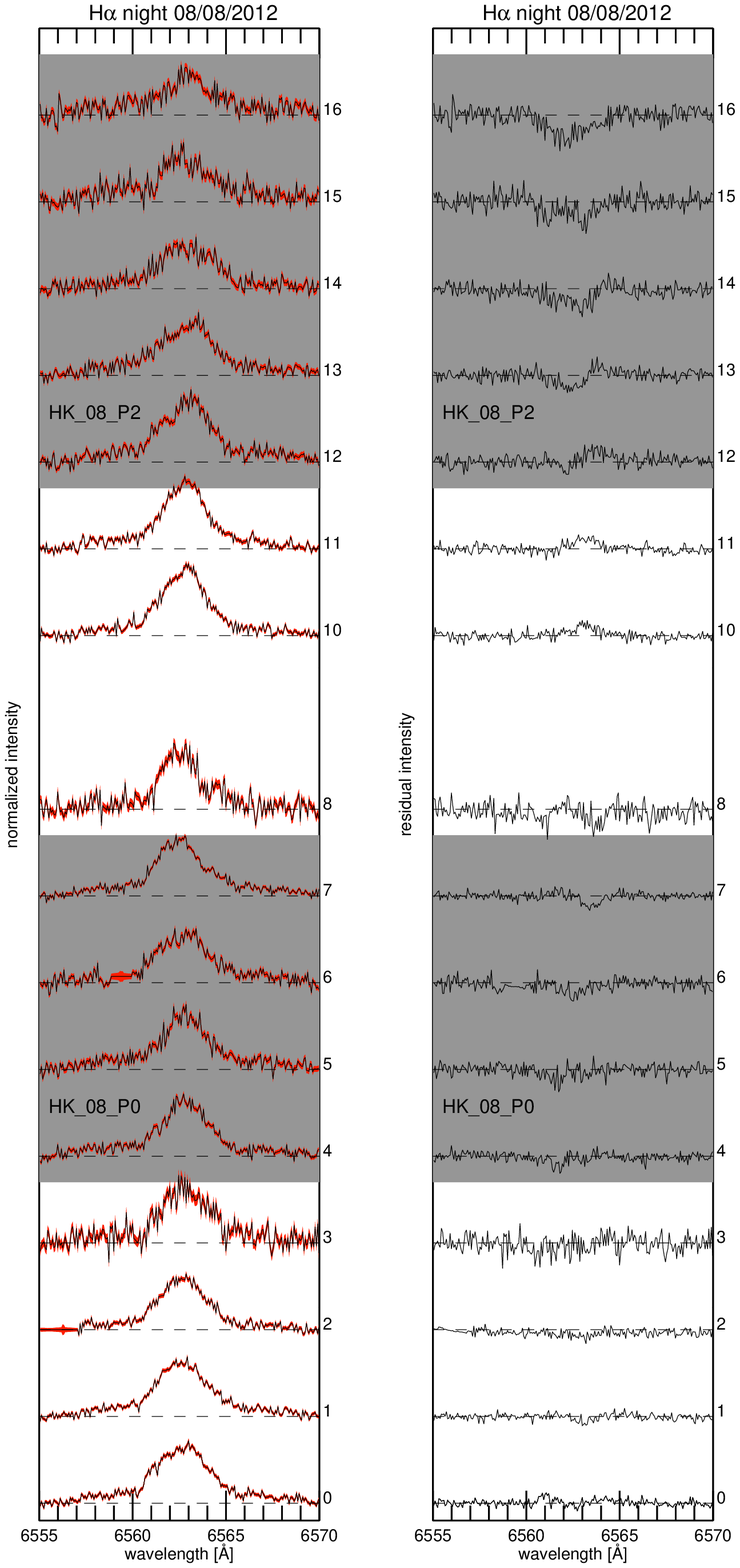}
 \caption{Left panel: Spectral H$\alpha$ time series of night 08/08/2012 of HK Aqr. Right panel: Residual H$\alpha$ time series of night 08/08/2012 of HK Aqr. Grey shaded area indicate the prominence spectra or residuals for that night.}
\label{allspec08}
 \end{center}
 \end{figure}
 
 \onecolumn

\begin{figure}
\begin{center}
\vspace*{0cm}
\includegraphics[width=10.0cm]{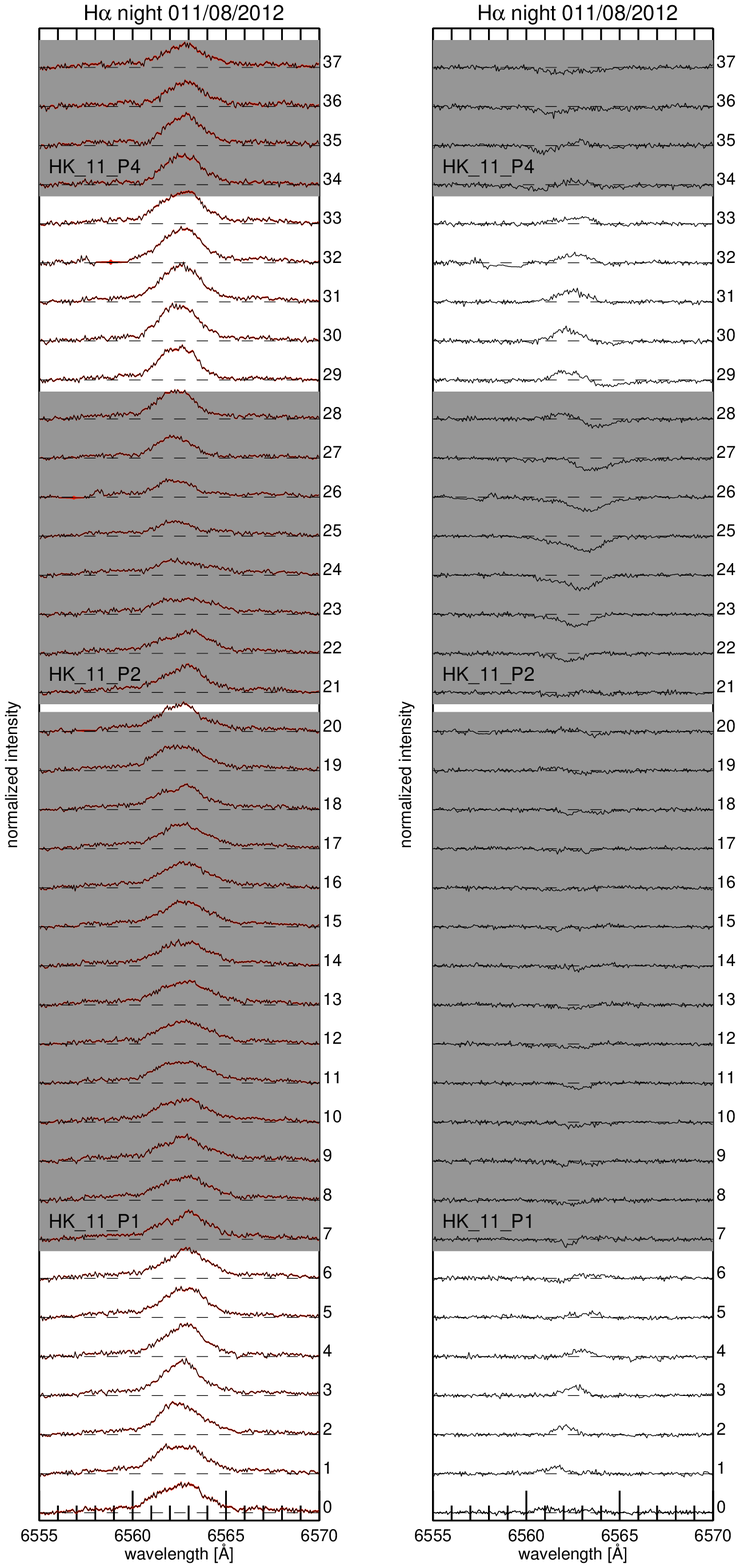}
 \caption{Left panel: Spectral H$\alpha$ time series of night 11/08/2012 of HK Aqr. Right panel: Residual H$\alpha$ time series of night 11/08/2012 of HK Aqr. Grey shaded area indicate the prominence spectra or residuals for that night.}
\label{allspec11}
 \end{center}
 \end{figure}
 
 \onecolumn

\begin{figure}
\begin{center}
\vspace*{0cm}
\includegraphics[width=10.0cm]{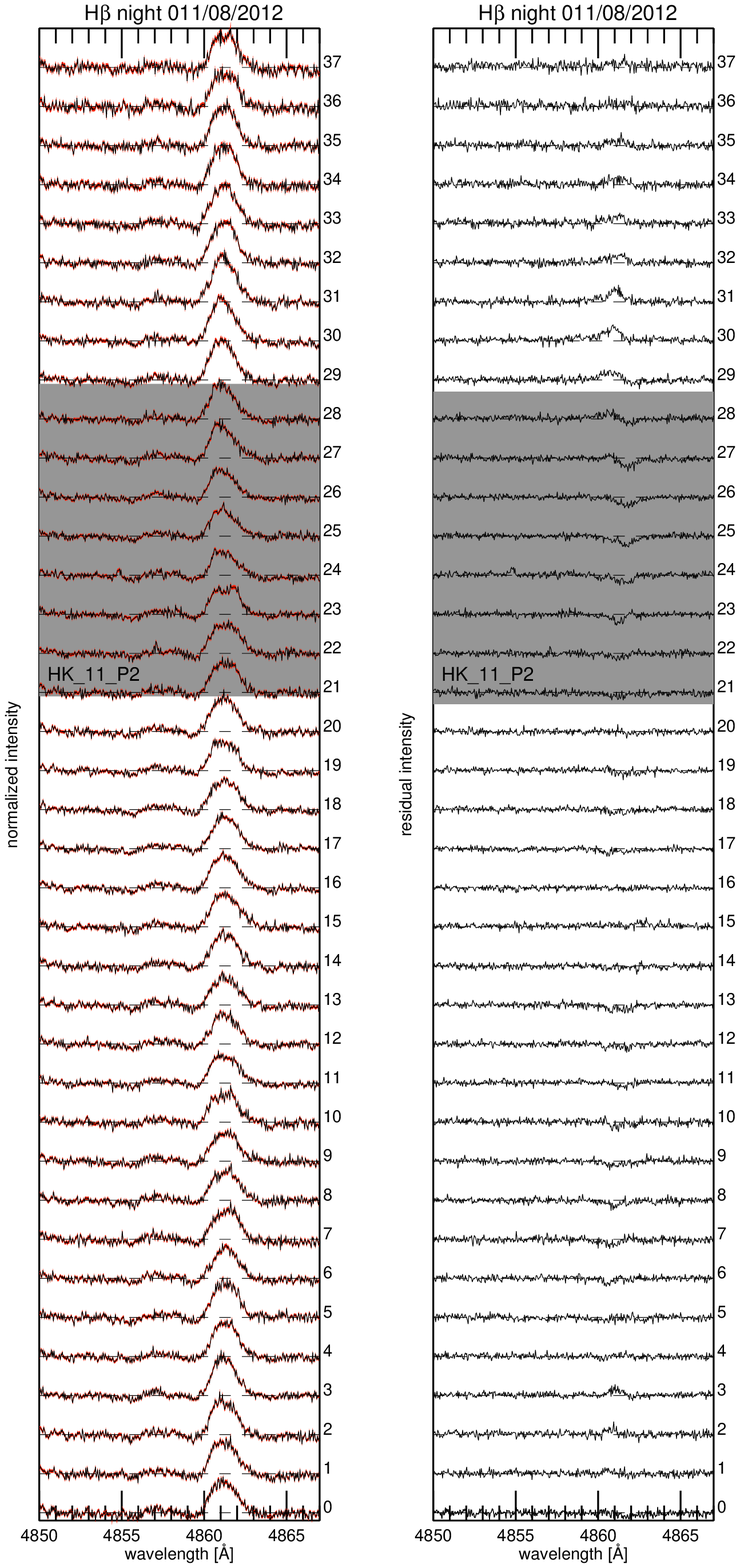}
 \caption{Left panel: Spectral H$\beta$ time series of night 11/08/2012 of HK Aqr. Right panel: Residual H$\beta$ time series of night 11/08/2012 of HK Aqr. Grey shaded area indicate the prominence spectra or residuals for that night.}
\label{allspec11beta}
 \end{center}
 \end{figure}

\onecolumn

\begin{figure}
\begin{center}
\vspace*{0cm}
\includegraphics[width=10.0cm]{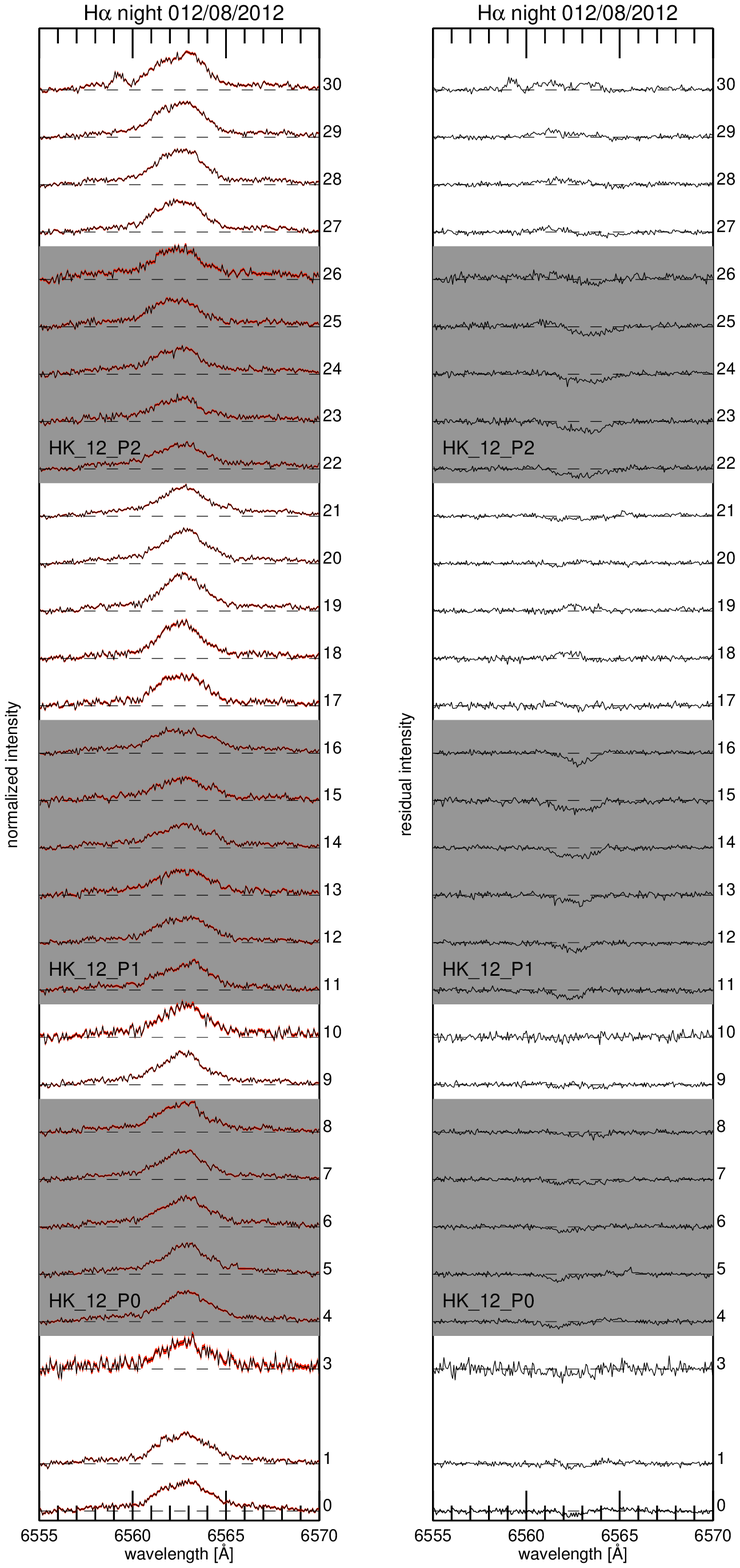}
 \caption{Left panel: Spectral H$\alpha$ time series of night 12/08/2012 of HK Aqr. Right panel: Residual H$\alpha$ time series of night 12/08/2012 of HK Aqr. Grey shaded area indicate the prominence spectra or residuals for that night.}
\label{allspec12}
 \end{center}
 \end{figure}
 
\onecolumn
 

\bsp	
\label{lastpage}

\end{document}